**Evidence of quantum vortex fluid in the mixed state of a very weakly pinned a-MoGe thin film**

Surajit Dutta, Indranil Roy, Soumyajit Mandal, Somak Basistha, John Jesudasan, Vivas Bagwe and Pratap Raychaudhuri[a]

*Tata Institute of Fundamental Research, Homi Bhabha Road, Colaba, Mumbai 400005, India.*

Quantum fluids refer to a class of systems that remain in fluid state down to absolute zero temperature. In this letter, using a combination of magnetotransport and scanning tunneling spectroscopy down to 300 mK, we show that vortices in a very weakly pinned *a*-MoGe thin film can form a quantum vortex fluid. Under the application of a magnetic field perpendicular to the plane of the film, the vortex state transforms from a vortex solid to a hexatic vortex fluid and eventually to an isotropic vortex liquid. The fact that the two latter states remain fluid down to absolute zero temperature is evidenced from the electrical resistance which saturates to a finite value at low temperatures. Furthermore, scanning tunneling spectroscopy measurements reveal a soft gap at the center of each vortex, which arises from large zero point fluctuation of vortices.

[a] pratap@tifr.res.in

In a Type II superconductor magnetic field penetrates inside the sample above the lower critical field ($H_{c1}$) in the form of quantized flux tubes called vortices, forming a periodic hexagonal lattice under their mutual interactions known as the Abrikosov vortex lattice[1,2] (VL). However, this periodic order is subject to several destabilizing factors. For example, the random pinning potential caused by crystalline defects in the superconductor can compete with the elastic energy of the VL and produce a disordered vortex glass[3,4,5,6,7]. Thermal fluctuations is another factor that can melt the VL into a vortex liquid[8,9]. While thermal fluctuations alone is often not enough to produce a vortex liquid in a bulk conventional superconductor[10,11], its effect can be greatly amplified in quasi 2-dimensional systems such as thin films or superconductors with strong anisotropy. Experimentally, signatures of vortex liquid have been obtained from magnetic and transport measurements in a number of quasi 2D systems such as weakly pinned superconducting thin films[12,13], layered High-$T_c$ cuprates[14,15,16] and organic superconductors[17].

In principle, quantum fluctuations can also destabilize a periodic solid. Well-known examples are $^3$He and $^4$He at ambient pressure, which remain liquid down to absolute zero temperature due to zero point fluctuations of the atoms. Quantum melting of the VL, has also been theoretically explored[18,19,20,21]. It has been suggested that when the magnetic field is increased, the zero point fluctuation amplitude of the vortices can exceed the Lindemann criterion[22] at which the VL melts into a vortex liquid. A dissipative state below the upper critical field, $H_{c2}$, that extends down to very low temperatures observed in several low dimensional superconductors has sometimes been attributed to the quantum melting of the vortices[23,24,25,26,27,28,29]. However, alternative explanations[30,31,32] have also been proposed.

Recently, using a combination of magnetotransport and scanning tunneling spectroscopy (STS) imaging[33] in a very weakly pinned thin film of the amorphous conventional superconductor MoGe (*a*-MoGe), we showed that vortex fluid states span a large region of the *H*-*T* parameter space. It was seen that the melting proceeds in two steps as the magnetic field is increased: First, the vortex solid (Bragg glass) melts into a hexatic vortex fluid (HVF), which retains the quasi-long-range orientational order at fields as low as $0.02H_{c2}$, followed by a transformation into an isotropic vortex liquid (IVL) above $0.8H_{c2}$. In this letter, we extend these measurements to much lower temperatures and show that the hexatic vortex fluid and the isotropic vortex liquid exist down to very low temperatures. Furthermore, detailed STS measurements on the vortex cores reveal the existence of large quantum zero point fluctuations of vortices that could be responsible for creating quantum fluid of vortices down to $T = 0$.

The samples used in this study consist of 20 nm thick *a*-MoGe films, similar to the one used in ref. 33. As shown before the films have extremely weak pinning as evidenced from a very low depinning frequency and the absence of any difference between the field cooled and zero field cooled states[33]. The electrical transport measurements were done using conventional 4-probe method in conventional $^3$He cryostat on a film capped with a protective Si layer to prevent oxidation. STS measurements were performed using a home-made scanning tunneling microscope[34] operating down to 350 mK. For STS measurements, post deposition, the film was transferred in the scanning tunneling microscope using an ultrahigh vacuum suitcase without exposure to air. Further details of samples and measurement are given in the supplementary document[35].

First, we present evidence that vortex fluid states exist down to very low temperatures. To identify the different vortex states we apply the same diagnostics as in ref. 33, but extending the measurements down to 300 mK. In a vortex solid[36] a few pinning centers can pin the entire VL thereby giving exponentially vanishing resistance below the flux flow critical current, $I_c$. In contrast, in a vortex fluid state the vortices are diffusive; consequently a finite linear resistance[37] appears even in the limit $I \rightarrow 0$. In Fig. 1(a) we show the linear resistivity, $\rho_{lin}$ at 450 mK measured from the linear slope of the *I-V* characteristics below 50 μA ( $\ll I_c$ ) measured as a function of magnetic field. At this temperature, $I_c > 300$ μA till about 95 kOe. The transition from the vortex solid to HVF occurs at the first characteristic field, $H_M^1 \sim 5$ kOe, where $\rho_{lin}$ becomes finite. At a higher field $H_M^2 \sim 70$ kOe, $\rho_{lin}$ starts rapidly increasing after going through a broad minimum and eventually reaches the normal state value $\rho_n$ above $H_{c2}$. This marks the transition from a HVF to IVL. The identification of these states is further confirmed from STS imaging[35] of the vortex state at 450 mK. The broad minima in $\rho_{lin}$ reflects a corresponding maxima in $I_c$ which arises at the boundary between HVF and IVL due to the shrinking of the topological defect free regions[33]. Tracking $H_M^1$ and $H_M^2$ from $\rho_{lin}$-$H$ at different temperatures (upper insets Fig. 1(a)) we construct the $H$-$T$ phase diagram of the vortex state (lower inset Fig. 1(a)). Both the $H_M^1$ and $H_M^2$ remain much smaller than $H_{c2}$ down to 300 mK ( ~ $0.04T_c$ ), in sharp contrast with the expectation for a thermally driven transition, where the melting line should approach[38] $H_{c2}$ in the limit $T \rightarrow 0$. To further confirm whether the HVF and IVL extend down to $T = 0$ we measure $\rho_{lin}$ as a function of temperature at various fields. Normally, for a current drive $I << I_c$, dissipation occurs in the mixed state due to thermally activated flux flow (TAFF) over the average pinning barrier, $U$, giving a temperature variation of the resistivity, $\rho = \rho_{ff}\exp[-U/kT]$, where $\rho_{ff}$ is the Bardeen-Stephen flux flow resistivity and $k$ is the Boltzmann constant. In a vortex solid (Bragg Glass)[36], $U(I) \propto \left(\frac{I_c}{I}\right)^{\gamma}$ (where $\gamma$ is a constant of the order of unity) and therefore $\rho$ exponentially drops to an immeasurably small value for currents smaller than $I_c$. In contrast, in a vortex fluid $U$ is independent of current giving a finite linear resistivity, which is expected to exponentially drop to zero with decrease in temperature. In Fig. 1(b), we plot $\rho$ -$T$ measured with a drive current of 50 μA. Fig. 1(c) shows the same data plotted in semi-log scale as a function of $1/T$. For $H \geq 10$ kOe, at low temperatures $\rho$ deviates from the expected activated behavior and saturates towards a constant value. The asymptotic finite value of $\rho$ as $T \rightarrow 0$ suggests that the vortices remain mobile down to absolute zero, as would happen if the vortices form a quantum fluid. In this context, it was recently pointed out that large current drives or external electromagnetic radiation could also give rise to resistance saturation due to electron overheating or vortex depinning[39]. Resistance saturation caused by large current drives should also be associated with strong non-linearity in the $I$-$V$ characteristics that we do not observe here[35]. To take care of external electromagnetic radiation, 1 MHz low-pass filters are connected at input and output stages of the cryostat[35]. Further filtering do not cause any further change in the $\rho$ -$T$ characteristics.

We now turn our attention to the spectroscopy of the vortex cores. In STS measurements the tunneling conductance, $G(V) = \frac{dI}{dV}\Big|_V$ between the tip and the sample gives the local tunneling density of states (LTDOS) at energy $eV$ ($e$ is the electron charge) with respect to Fermi energy. In a conventional superconductor the vortex consists of circulating supercurrents with a normal metal core with radius of the order of the coherence length, $\xi$, at the center. Consequently, while far from the vortex core the LTDOS exhibits a superconducting gap and coherence peak partially broadened due to the orbital current, inside the core the LTDOS is either flat or exhibits a small peak at zero bias due to the formation of Caroli–De Gennes–Matricon[40] bound states in very clean samples. Here, we capture the full $G(V)$ vs. $V$ spectroscopic map at 450 mK over a small area containing 6-8 vortices (Fig. 2(a)-(b)). Fig. 2(c) and 2(d) show the normalized conductance spectra, $G_N(V)$ vs. $V$, (where, $G_N(V) = \frac{G(V)}{G(V=4\ mV \gg \Delta/e)}$ and $\Delta \sim 1.3$ meV is the superconducting energy gap) along the line passing through the center of a vortex core at two representative fields. Though the superconducting gap starts to fill while approaching the vortex center, a soft gap in the LTDOS continues to exist even at the center of the vortex (i.e. $G_N(0) < 1$). In Fig 2(e) and 2(f) we show the variation of $G_N(0)$ along three lines passing through the vortex center. In Fig. 2(g) we plot $G_N(0)$ at the center of the vortex and at the midpoint between two vortices as a function of $H$. Here, each point is obtained by taking the average value of $G_N(0)$ at 6 vortex centers. We observe that $G_N(0)$ shows an overall increasing trend with magnetic field except for two anomalies around 5 kOe and 70 kOe.

A soft gap in the vortex core has earlier been observed in unconventional high-$T_c$ cuprate superconductors[41,42] and in strongly disordered[43,44] NbN thin films. It has been attributed alternatively to the presence of phase incoherent Cooper pairs[41,44] in the normal state, or to a competing order[42] (in cuprates). In *a*-MoGe, which is a conventional superconductor closely following Bardeen-Cooper-Schrieffer (BCS) theory, these explanations seem unlikely. Furthermore, STS maps do not reveal any significant inhomogeneity as in NbN, which could render the system susceptible to phase fluctuations. On the other hand, the soft gap can be understood if we assume that vortex core spatially fluctuates rapidly about its mean position. Since STS is a slow measurement, such rapid fluctuation (superposed on the slow diffusive motion in the HVF and IVL states) will get integrated out, showing only the average tunneling conductance at a particular location. The qualitative effect of this fluctuation is that the tunneling conductance close to the center of the vortex would have contribution both from the vortex core as well as regions outside it.

We now make this argument more quantitative. The exact calculation of the LTDOS in the presence of vortices is in-principle possible by solving the Usadel equations[45], but this computation is very difficult[46]. Here, instead we adopt a phenomenological approach. First we simulate the regular VL in a superconductor where the vortices are static. We assume that at the vortex core of an isolated vortex $G_N(V) = 1$, whereas far from the core, $G_N(V) = G_N^{BCS+\Gamma}(T,V)$ given by superconducting density of states obtained from BCS theory[47]. $G_N^{BCS+\Gamma}(T,V)$ is obtained by fitting the experimental zero field tunneling conductance spectra at the same temperature, with $\Delta$, and a broadening parameter[48], $\Gamma$, used as fitting parameters. To interpolate between the two we use an empirical Gaussian weight factor, $f(\boldsymbol{r}) = \exp[-\frac{r^2}{2\sigma^2}]$, such that $G_N(V,\boldsymbol{r}) = f(\boldsymbol{r}) + [1 - f(\boldsymbol{r})]G_N^{BCS+\Gamma}(T,V)$. For a VL, the resultant normalized conductance, $\tilde{G}_N(V,\boldsymbol{r})$, is assumed to be a linear superposition of the conductance from all vortices, i.e. $\tilde{G}_N(V,\boldsymbol{r}) = \sum_i G_N(V,\boldsymbol{r}-\boldsymbol{r_i}) / [\sum_i G_N(V=0,\boldsymbol{r}-\boldsymbol{r_i})]_{max}$, where $\boldsymbol{r}_i$ is the position of the $i$-th vortex and sum runs over all vortices; the normalizing factor ensures that the $\tilde{G}_N(V=0) = 1$ at the center of each vortex. By taking, $\sigma \approx 1.17\xi$, we cross-checked that this procedure reproduces well the experimental tunneling conductance in the VL in a $NbSe_2$ single crystal[35].

To simulate the situation when the vortices are randomly fluctuating about their mean positions, we calculate $\tilde{G}_N(V,\boldsymbol{r})$ for 200 realizations of a distorted hexagonal lattice (lattice constant $a$), where each lattice point is displaced by a random vector, $\delta\boldsymbol{r}_i$, satisfying the constraint, $|\delta\boldsymbol{r_i}| \leq \alpha < a$, and compute the average $\langle\tilde{G}_N(V,\boldsymbol{r})\rangle_\alpha$. $\alpha$ is the amplitude of fluctuations. Fig. 3(a) and 3(b) show representative conductance maps simulated for a VL at 10 kOe for $V = 0$ without and with incorporating the fluctuation of the vortices. Fig. 3(c) and 3(d) show the corresponding conductance maps for $V = 1.45\ mV$, close to the coherence peaks.

To fit the experimental data we take $\alpha$ and $\Gamma$ as fitting parameters and constrain $\xi$ between 3.9-4.5 nm, which is within 10% of the value obtained from $H_{c2}$ ($\xi \sim 4.3$ nm). Fig. 3(e)-(g) show the line-cuts of $G_N(0,r)$ along with $\langle \tilde{G}_N(V,r) \rangle_\alpha$ at 3 different fields. Above 10 kOe, we use a larger $\Gamma$ (compared to its zero field value) to account for the additional broadening from the orbital current around vortices. For $V = 1.45$ $mV$, the same set of parameters reproduces the qualitative variation of $G_N(V = 1.45\ mV, r)$, but the conductance value is overestimated by 15-20%; this is most likely because the phenomenological $\Gamma$ parameter does not account for the suppression of the coherence peak due to orbital currents accurately.

To understand the physical origin of this fluctuation we note that vortices, separated by distance $R$, exert a repulsive force[49] $F_v = \frac{\Phi_0^2 d}{2\pi\mu_0\lambda^2}\left(\frac{1}{R}\right)$ on each other; here $\Phi_0$ is the flux quantum, $\lambda$ is the London penetration depth and $\mu_0$ is the vacuum permeability and $d$ is the film thickness. Thus each vortex is confined in a potential well formed by the repulsion from surrounding vortices. Since the vortex motion is overdamped we assume that each vortex oscillates individually in this potential well. When the oscillation amplitude, $\alpha \ll a$, the potential can be approximated to a harmonic oscillator with effective spring constant[35], $K = \frac{\Phi_0^2 d}{16\pi\mu_0\lambda^2}\left(\frac{1}{a^2}\right) \equiv \frac{K_0}{16a^2}$. This picture remains valid even in the vortex fluid states as long as the diffusive motion of vortices remains much slower compared to the fluctuation frequency. The total energy of each vortex in this potential well is given by $m_v\omega^2\alpha^2$, where $m_v$ is the vortex mass and $\omega^2 = K/m_v$. The amplitude of zero point motion is obtained by equating this to the zero point energy ($\hbar\omega$) of the quantum harmonic oscillator, giving, $\frac{\alpha}{a} \sim \frac{2\hbar^{1/2}}{(K_0 m_v)^{1/4}}\left(\frac{1}{a}\right)^{1/2}$. Since, $a \propto H^{-1/2}$, this implies $\frac{\alpha}{a} \propto H^{1/4}$. In contrast, if the oscillation is thermal in origin, $m_v\omega^2\alpha^2 \sim kT$. This gives, $\frac{\alpha}{a} \sim \left(\frac{kT}{K_0 m_v}\right)^{1/2}$, which is independent of magnetic field.

To test this, in Fig. 3(h) we plot the magnetic field variation of $\alpha/a$ at 450 mK. Here $\alpha$ is obtained by fitting $G_N(0, r)$ profile and further averaging over 6 vortices at each magnetic field. $\alpha$ shows an overall increase with $H$ and two anomalies close to 5 kOe and 70 kOe which we discuss later. The increasing trend is well captured by $\alpha/a \propto H^{1/4}$, which clearly rules out thermal fluctuations. From the coefficient of the fit and using the experimental value of $\lambda \approx 534\ nm$ we obtain $m_v \sim 35 m_e$ , where $m_e$ is the electron mass. Though different estimates of $m_v$ considerably vary[50,51,52,53,54,55] we can compare this value with the most widely used estimate[49] , $m_v = \frac{2}{\pi^3} m^* k_F d$, where $m^*$ is the effective mass and $k_F$ is the Fermi wave vector. Assuming $m^* = m_e$, and using the free electron expression, $k_F = (3\pi^2 n)^{1/3}$, where $n = 5.2*10^{29}$ el/m$^3$ is the carrier density determined from Hall effect[35], we obtain $m_v \approx 32 m_e$ which is consistent with the value obtained from our experiments. Coming to the cusp-like anomalies observed at 5 kOe and 70 kOe we note that these anomalies appear very close to the vortex solid to HVF, and HVF to IVL boundaries respectively. These anomalies most likely arise from the anharmonicity of the confining potential close to the phase boundaries, though a detailed understanding will require further investigations. It is interesting to note that the first anomaly appears when $\frac{\alpha}{a} \sim 0.14$, which is consistent with the Lindemann criterion for melting.

In summary, the existence of quantum vortex liquid and quantum zero point motion in a very weakly pinned superconducting thin film of *a*-MoGe presents a physical scenario under which we could realize a state with saturating resistance extending down to $T = 0$ in a superconductor below $H_{c2}$, similar to that in Bose metals and failed superconductors reported in a number of superconducting thin films in the presence of magnetic field. Furthermore, if the vortices indeed remain a fluid down to $T = 0$, one cannot preclude the possibility that they will condense in a superfluid state at a lower temperatures, producing a “super-insulating” state[56]. This possibility should be explored by extending these measurements to lower temperatures.

*Acknowledgements:* We would like to thank Premala Chandra and her graduate solid state class, Lara Benfatto, Dragana Popovic, Ilaria Maccari and T. V. Ramakrishnan for reading the manuscript and providing valuable suggestions. We would like to thank Rajdeep Sensarma, Kedar Damle, Chandan Dasgupta, C. Brun and Steve Kivelson for valuable discussion. We thank Arumugam Thamizhavel for his help in preparing the MoGe target. We thank Rudheer Bapat and B. A. Chalke for TEM characterization of the samples. The work was supported by Department of Atomic Energy, Govt. of India and Department of Science and Technology, Govt. of India (Grant Nos. EMR/2015/000083 and INT/Italy/P-21/2016(SP)).

IR and SD performed the STS measurements and analyzed the data. SD, SB, SM performed the transport and magnetic measurements and analyzed the data. JJ and VB optimized deposition conditions and synthesized the samples. The theoretical analysis were carried out by SD and IR in consultation with PR. PR conceived the problem, supervised the experiments and wrote the letter. All authors read the manuscript and commented on the letter.

---

[50] H. Suhl, *Inertial Mass of a Moving Fluxoid*, Phys. Rev. Lett. **14**, 226 (1965).

[51] E. M. Chudnovsky and A. B. Kuklov, *Inertial Mass of the Abrikosov Vortex,* Phys. Rev. Lett. **91,** 067004 (2003).

[52] N. B. Kopnin and V. M. Vinokur, *Dynamic Vortex Mass in Clean Fermi Superfluids and Superconductors*, Phys. Rev. Lett. **81,** 3952 (1998).

[53] M. W. Coffey, *On the inertial mass of a vortex in high-$T_c$ superconductors: closed form results*, J. Phys. A: Math. Gen. **31**, 6103 (1998)

[54] E.Šimánek, *Inertial mass of a fluxon in a deformable superconductor*, Phys. Lett. A **154,** 309 (1991).

[55] D. M. Gaitonde and T. V. Ramakrishnan, *Inertial mass of a vortex in cuprate superconductors*, Phys. Rev. B **56**, 11951 (1997).

[56] V. M. Vinokur, T. I. Baturina, M. V. Fistul, A. Yu. Mironov, M. R. Baklanov and C. Strunk, *Superinsulator and quantum synchronization,* Nature **452,** 613 (2008).

**Figure Captions**

**Figure 1**| (a) Variation of $\rho_{lin}$ as a function of magnetic field at 450 mK. $H_M^1$ marks the transition from vortex solid to HVF and $H_M^2$ from HVF to IVL. *Upper insets* show $\rho_{lin}$ vs *H* at different temperatures close to the phase boundaries between vortex solid-HVF (left) and HVF-IVL (right). Every successive curve is shifted by a vertical offset 0.015, except the curve at 0.3 K. *The lower inset* shows the phase diagram in the *H-T* parameter space; $H_M^1$ is multiplied by a factor of 2 on magnetic field axis for clarity. (b) Plot of $\rho$ vs *T* in different magnetic fields expanded in the low temperature regime. (c) $\rho$ vs. *1/T* in semi-log scale.

**Figure 2**| (a)- (b) VL image obtained from $G(V)$ maps at $V = 1.45\ mV$, at 450 mK for 5 kOe and 65 kOe respectively. The three dotted lines are passing through one vortex core. (c)- (d) Spectra along the red dotted lines of (a) and (b) respectively. The black (pink) dotted line denotes spectra at (away from) the vortex center which are shown in insets of (e) and (f). (e)- (f) Variation of $G_N(0)$ along three lines in (a) and (b) respectively. The grey transparent line denotes average $G_N(0)$ variation of these three lines. (g) Magnetic field variation of $G_N(0)$ values at the centers of (away from) vortex cores are shown by the connected red (green) squares. Two anomalous peaks in $G_N(0)$ variation at the vortex centers are observed close to 5 kOe and 70 kOe.

**Figure 3**| (a)- (b) Simulated zero bias conductance maps, $\langle \tilde{G}_N(0,r) \rangle$ for VL at 10 kOe without and with incorporating fluctuation of vortices respectively. (c)- (d) Corresponding conductance maps ($\langle \tilde{G}_N(V,r) \rangle$) for $V = 1.45\ mV$, denoted as $\langle \tilde{G}_{Np} \rangle$. (e)- (g) The red connected datapoints are averaged line-cuts of $G_N(0,r)$ along vortex cores for 1.75, 7 and 65 kOe respectively. The black lines are line-cuts of simulated $\langle \tilde{G}_N(0,r) \rangle_\alpha$ of the corresponding field. (*inset*) Line-cuts of $G_{Np}$ or $G_N(V = 1.45\ mV)$ along vortex cores (red lines) for the same fields along with line-cuts of simulated $\langle \tilde{G}_N(V = 1.45\ mV, r) \rangle_\alpha$ (black lines). (h) The black connected dots show the magnetic field variation of $\alpha/a$ at 450 mK, which is obtained by fitting the experimental $G_N(0,r)$ profiles. The red line is a fit to the $\alpha/a$ values, which goes as $H^{1/4}$. The anomalies in the variation of $\alpha/a$ are marked by two green arrows.

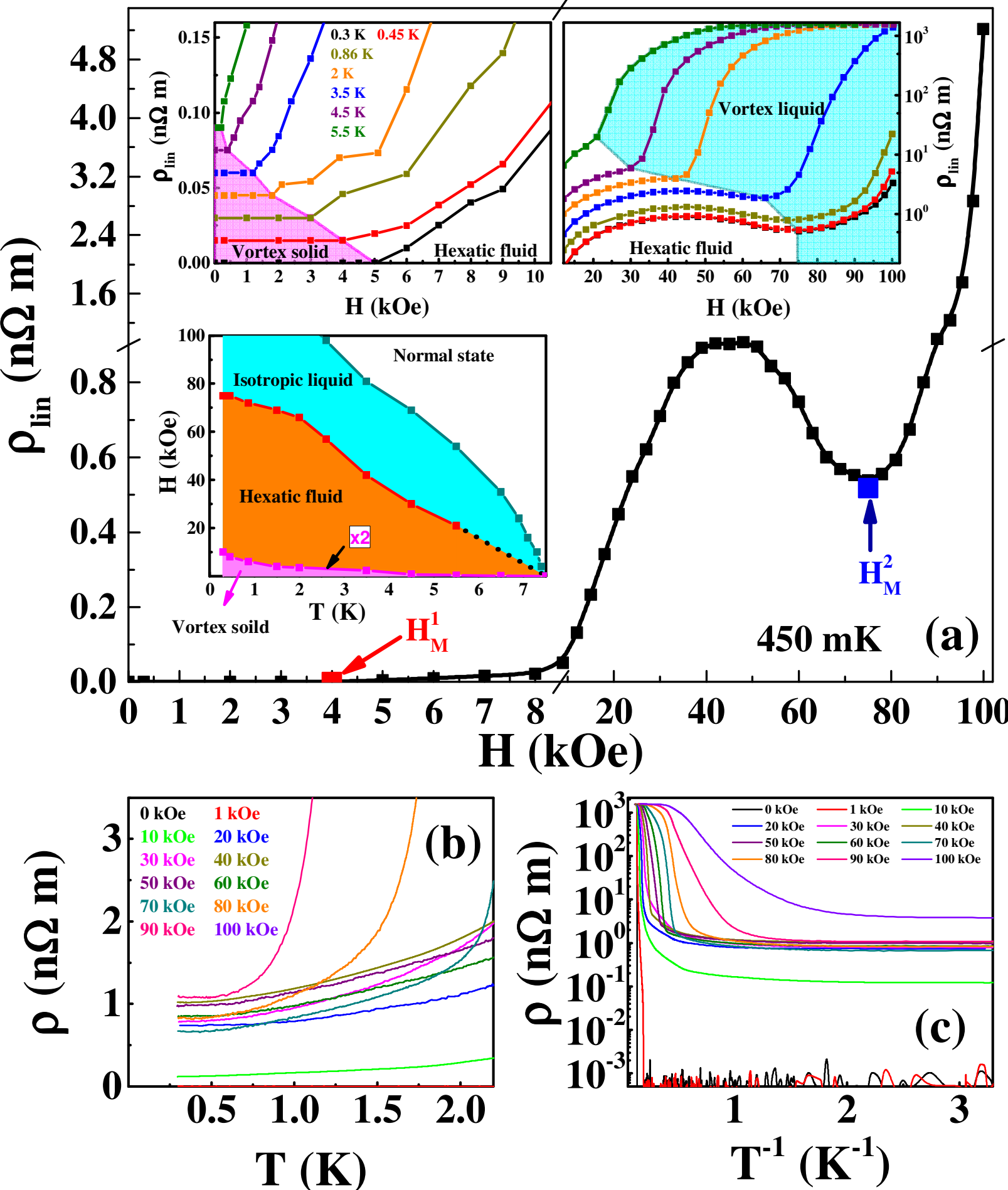

Figure 1

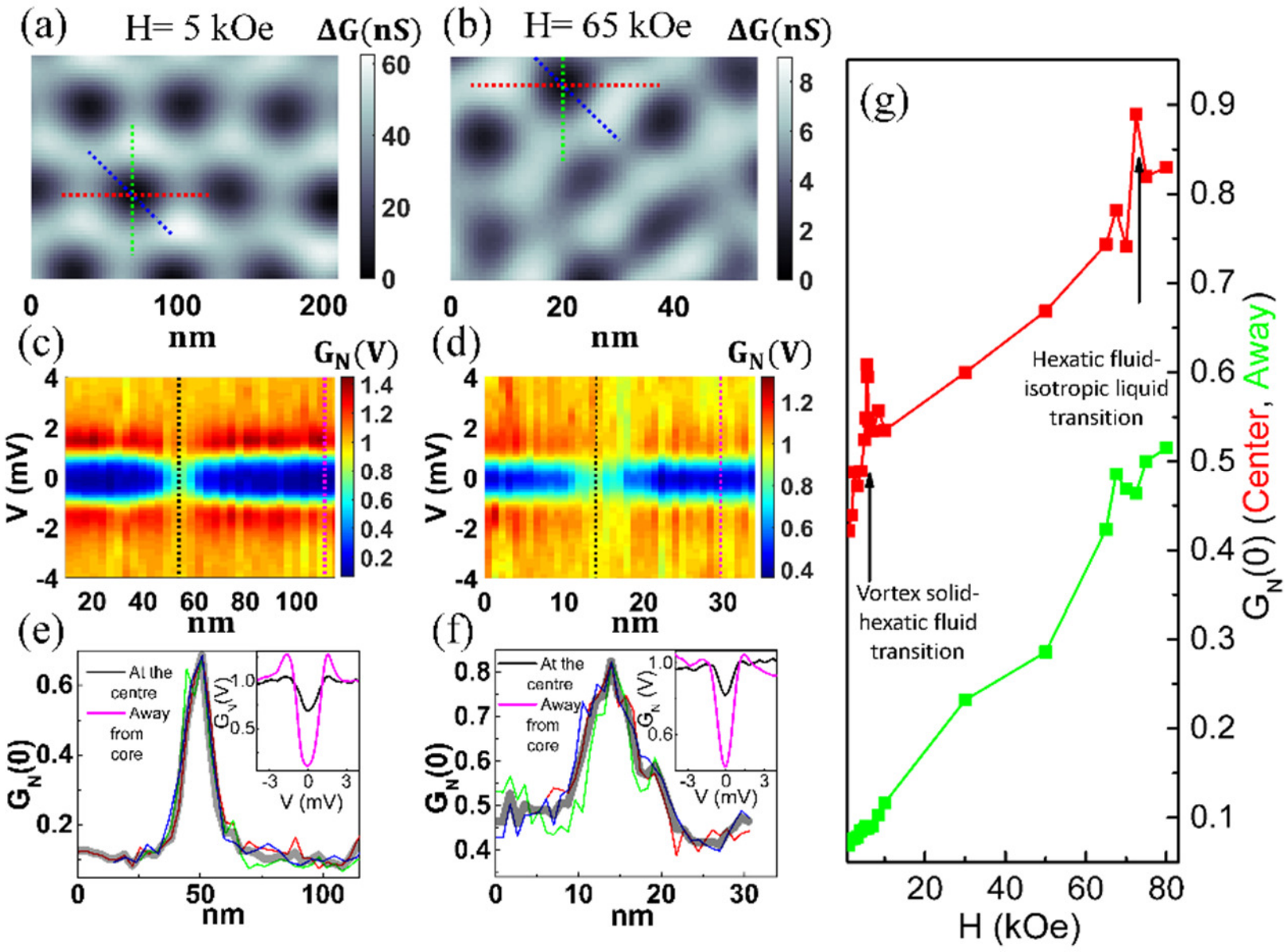

Figure 2

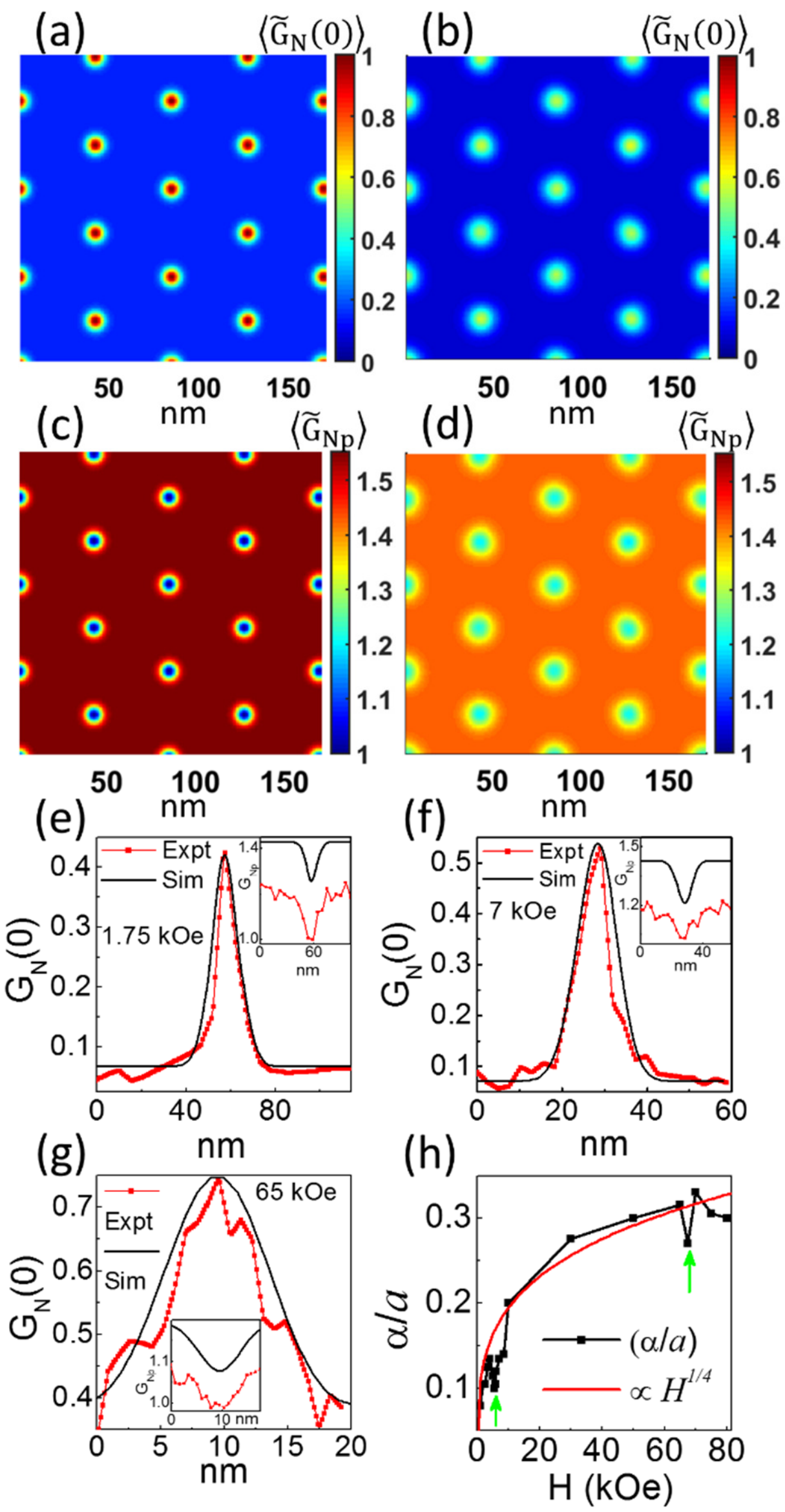

Figure 3

## Supplementary Material

### I. Sample growth, characterization and experimental details

***Sample growth.*** The *a*-MoGe thin films used in this study were grown on surface oxidized Si substrate through pulsed laser deposition. The thickness of the films, $d$ ~ 20 nm. The amorphous film was synthesized by ablating a $Mo_{70}Ge_{30}$ bulk target prepared by arc-melting Mo and Ge in stoichiometric proportions in Ar ambient. For laser ablation, we used a 248 nm excimer laser keeping the substrate at room temperature. The laser was focused in a tight spot with repetition rate of 10 Hz on the target, giving an effective energy density ~ 240 mJ/mm$^2$ per pulse. The deposition is carried out in vacuum of $1\times10^{-8}$ Torr. The growth rate was ~ 1 nm/100 pulse. The uniform amorphous nature of the sample was confirmed through high resolution transmission electron micrographs.

***Electrical measurements.*** Electrical measurements (resistivity and Hall coefficient) were performed using conventional 4-probe method using a d.c. current source and a nanovoltmeter in a $^3$He cryostat. The samples were deposited in the shape of a 6-terminal Hall bar as schematically shown in Fig. s1. One important aspect of the measurement setup was careful shielding from external electromagnetic radiation. For this low-pass RC filters with cutoff frequency of 1 MHz were connected on the electrical feedthrough to the sample. Without the filters, the resistance showed anomalously large saturation values at low temperatures as shown in Fig. s2. Incorporating additional filters did not cause any noticeable change in the data.

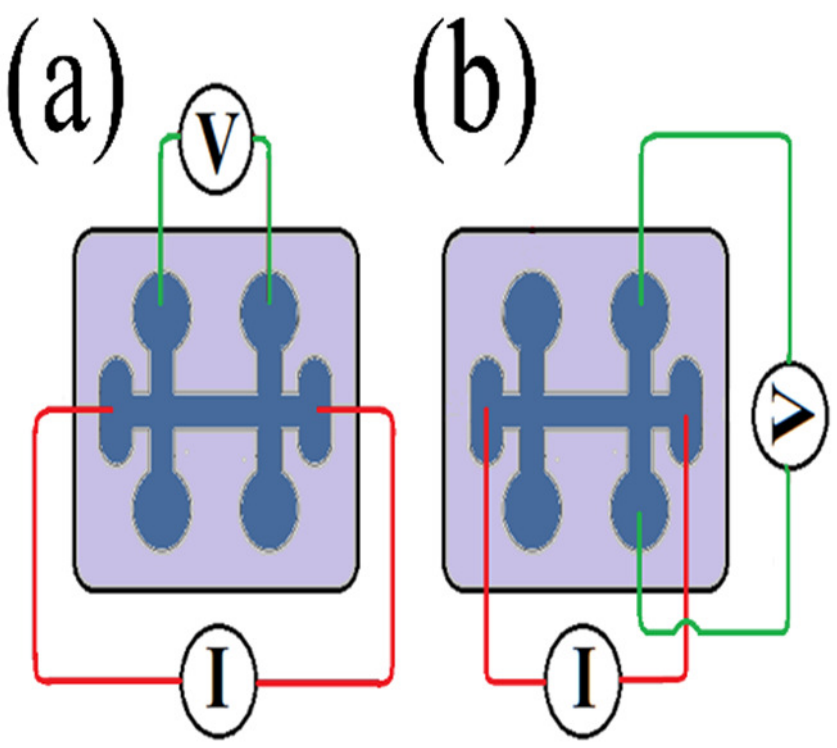

**Figure s1|** Schematic configuration of the Hall bar used for (a) resistivity and (b) Hall measurements. The width of the channel between the voltage probes is 300 μm.

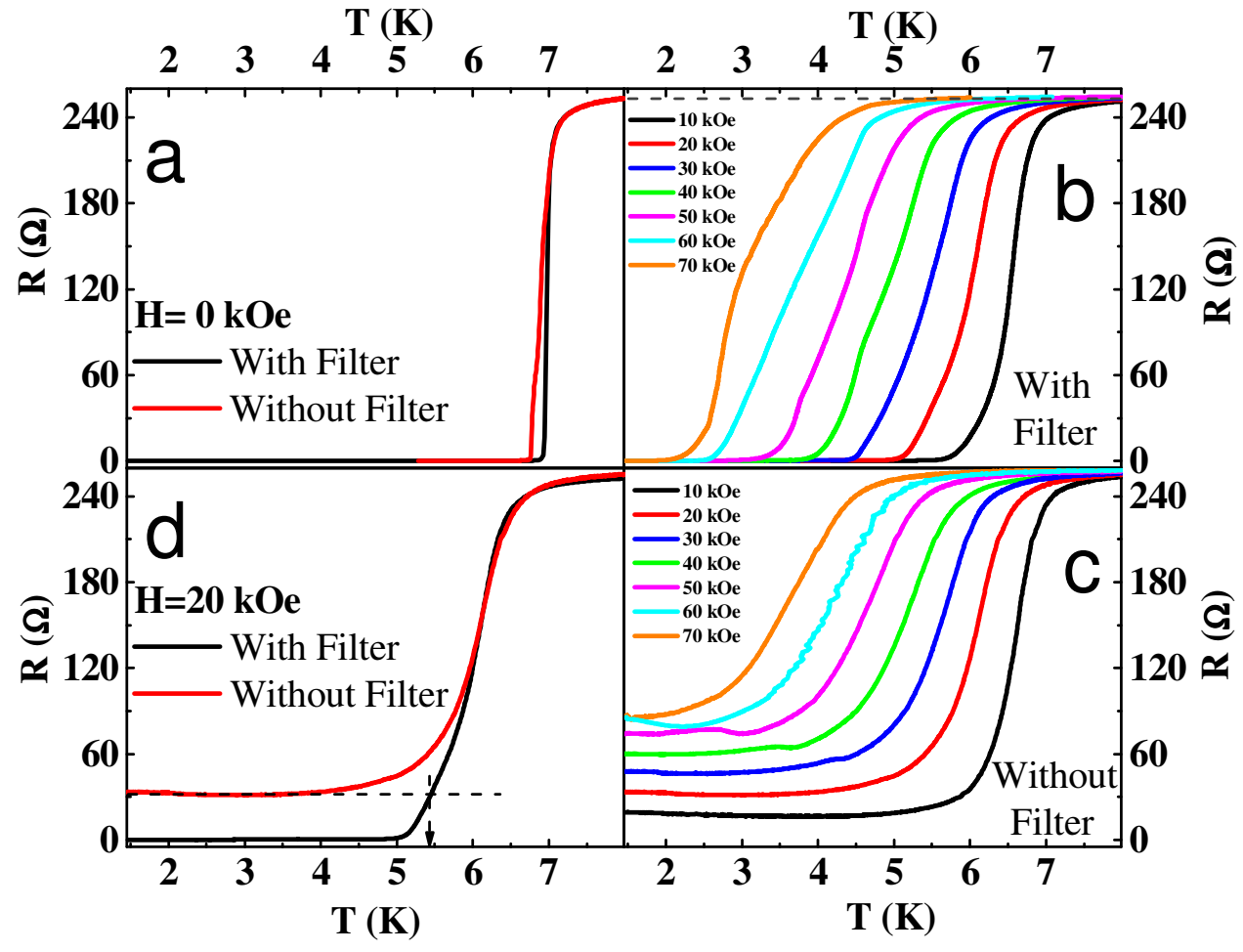

**Figure s2|** (a) Resistance vs Temperature (R-T) in zero field: With filter data shows a sharp superconducting transition at 7 K; without filter data shows a broadened transition at 6.7 K. (b) R-T data taken with filter for different magnetic fields: showing decrease in transition temperature with increasing field, but in all cases resistance drops to a very small value. (c) R-T data taken without filter for different magnetic fields showing a large saturating resistance at low temperatures. Magnitude of the saturating resistance increases with increasing magnetic field. (d) Comparison of with filter and without filter R-T data for 20 kOe: the dashed line shows saturating value of resistance in low temperature for the without filter data; Corresponding temperature required to obtain that resistance for with filter data is about 5.5 K.

***STS measurements.*** STS measurements were performed in a home-built scanning tunneling microscope (STM) operating down to 350 mK and fitted with 90 kOe superconducting solenoid. Measurements were performed using a normal metal Pt-Ir tip. The differential tunneling conductance, *G(V)* was measured by superposing a small a.c. voltage, $V_{ac}$ (150 μV, 2.11 kHz) on the d.c. bias voltage $V_{dc}$ and measuring the a.c. current $I_{ac}$ using a lock-in amplifier, so that, $G(V = V_{dc}) = dI/dV|_{V=V_{dc}} \approx I_{ac}/V_{ac}$. To image the VL, *G(V)* maps were recorded with the bias voltage close to the superconducting coherence peak, such that each vortex appears as a local minima in the conductance. The full *G(V)-V* spectra were recorded by switching off the feedback at a given location, and recording *G(V)* while the bias voltage is swept from positive to negative bias. The time to acquire a complete spectrum is ~ 2 seconds. While acquiring full spectroscopic area maps each spectra was averaged over three sweeps at every point.

***Characterization of Superconducting properties.*** The temperature variation of the penetration depth, λ, and superconducting energy gap, Δ, has been reported in the supplementary of ref. 1. The additional two parameters that were determined for the purpose of this work were the carrier concentration, *n*, and superconducting coherence length, ξ.

The *carrier concentration* was determined from the Hall coefficient measured at 25 K. The Hall voltage was determined from reversed field sweeps from 85 kOe to -85 kOe after subtracting the resistive contribution. Fig. s3 shows the transverse resistivity, $\rho_{xy}$, as a function of magnetic field, $H$; the Hall coefficient, $R_H = \frac{\rho_{xy}}{H} = \frac{1}{ne}$, (where $e$ is the electron charge) is obtained from the slope of this curve. We obtain $n = 5.2 * 10^{29}$ $el/m^3$.

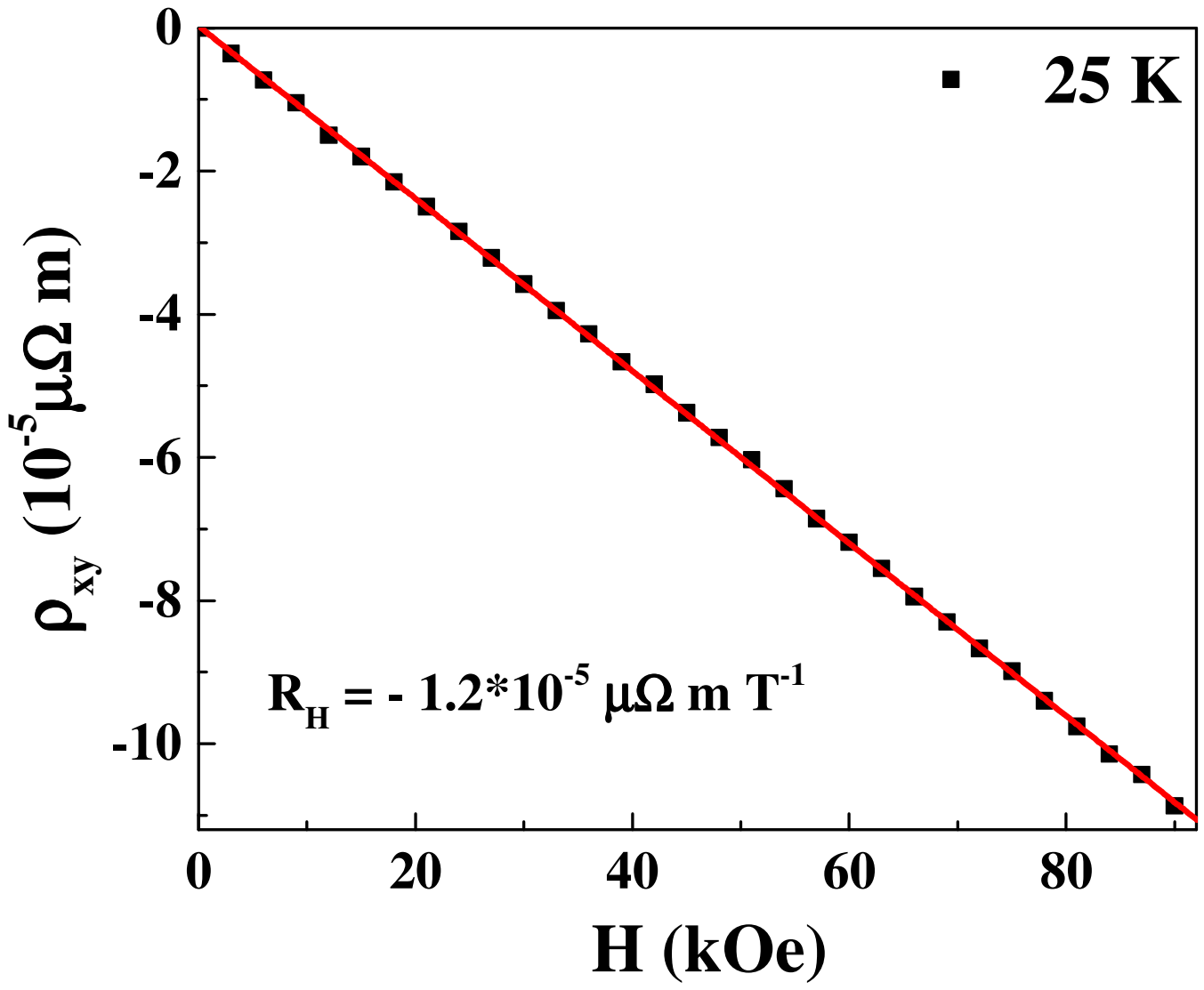

**Figure s3|** $\rho_{xy}$ as a function of $H$ at 25 K. The Hall coefficient $R_H$ is obtained from the slope of this curve.

The *coherence length* was determined from the temperature variation of the upper critical field close to $T_c$ using the dirty limit formula[2],

$$H_{c2}(0) = 0.69 \times \left(\frac{dH_{c2}}{dT}\right)_{T=T_c} \text{ and } \xi = \sqrt{\frac{\Phi_0}{2\pi H_{c2}(0)}}.$$

We obtain, $H_{c2}(0) \approx 180\ kOe$ corresponding to $\xi \sim 4.3\ nm$.

***Magnetic field and Temperature variation of the Flux Flow Critical current.*** The flux flow critical current, $I_c$, was determined from the *I-V* characteristics over the whole range of temperature and magnetic field. In principle, for a vortex solid at $T = 0$, $I_c$ corresponds to a hard threshold in the *I-V* curve where the voltage appears. In practice this sharp threshold is rounded off, and therefore $I_c$ has to be determined by extrapolating back the linear region of the *I-V* curve in the flux flow regime. This is illustrated for some representative temperatures and fields in Fig. s4 (left panels). In Fig. s4 (right panels), we show the magnetic field variation of $I_c$ at different temperatures. $I_c$ shows a shallow maxima at the boundary between HVF and IVL, the "peak effect", whose origin has been explained in detail in ref. 1. This peak in $I_c$ is reflected as a shallow minimum in $\rho$-$H$.

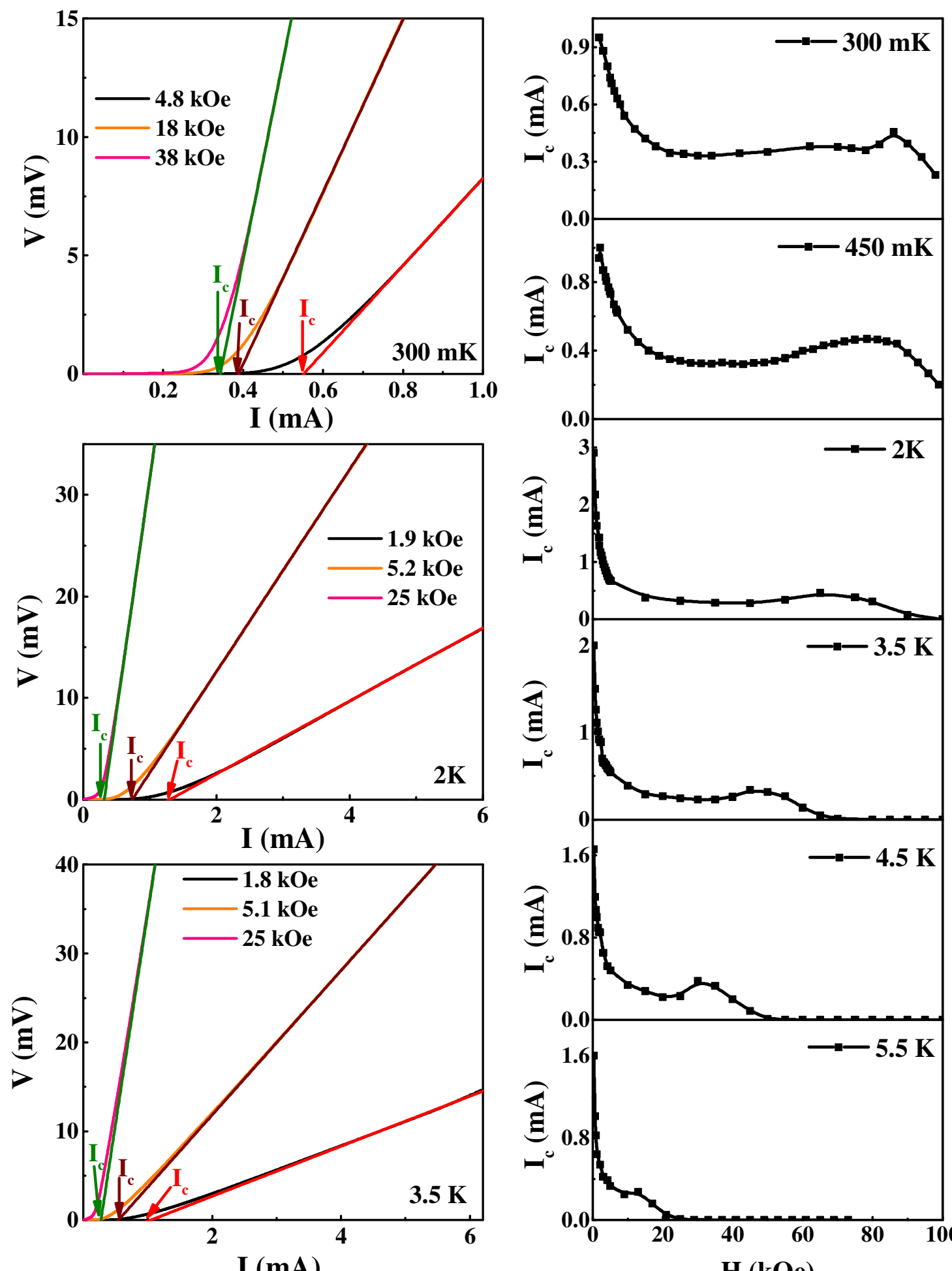

**Figure s4|** The left panels show representative *I-V* curves, along with linear fits to the flux flow regime at different temperatures and magnetic fields; $I_c$ is obtained from the intercept of the linear fits with the x-axis. The right panels show the variation of $I_c$ with magnetic field at different temperatures.

***I-V characteristics at 450 mK for I << $I_c$*.** Unlike in a vortex solid where the voltage exponentially drops to zero below $I_c$, a vortex fluid exhibits finite linear resistance even below $I_c$. Fig. s5 shows the *I-V* characteristics as a function of magnetic field at 450 mK, in the range $I$ = 0 - 100 μA. The nearly perfect linearity over this large range confirms the existence of vortex fluid state at this temperature.

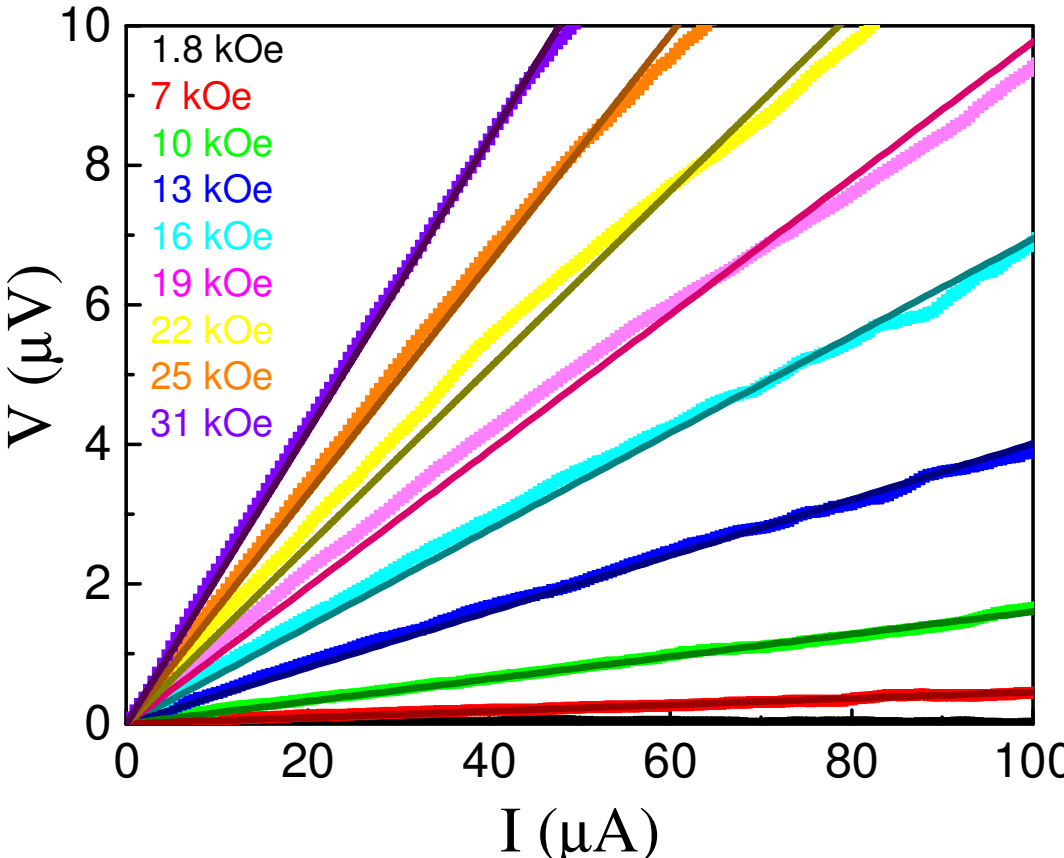

**Figure s5|** *I-V* characteristics at 450 mK for $I << I_c$. The solid lines are linear fit to the data.

**II. Identification of the Vortex Solid, HVF and IVL from STS imaging**

The vortex solid, HVF and IVL states can be unambiguously identified from the structure of the VL imaged using STS. In ref. 1, we had presented the structural evolution of the VL with magnetic field at 2 K. In Fig. s6 , we present VL images at 450 mK as a function of magnetic field. The position of each vortex is obtained from the local minima of the conductance map acquired at bias voltage close to the superconducting coherence peak and topological defects are identified through Delaunay triangulation. At 2.5 kOe and 4 kOe, we observe a hexagonal vortex lattice without any topological defects as expected for a vortex solid. Between 6 kOe and 70 kOe we see the gradual proliferation of dislocations (adjacent pair of points with 5-fold and 7-fold coordination) which drives the system into a HVF. These dislocations destroy the positional order but retains a quasi-long range orientational order of the VL. Consequently, the Bragg spots in the 2D Fourier spectrum get broadened but retain the six-fold symmetry. At 85 kOe, the orientational order is destroyed and the Fourier transform shows a diffuse ring, corresponding to an IVL.

To further confirm these identifications we also perform measurements of vortex creep, by capturing 10 to 12 consecutive vortex images at the same area at 15 minutes interval. The evolution of vortex creep with increasing magnetic field gives key insights to the nature of the VL. The vortices undergo small meandering motion about their mean position in the vortex solid phase (fig. s7 (a)-(b)) but do not show any large jump, whereas in the hexatic vortex fluid (HVF) phase the vortices move along one of the three principal axes of the VL (fig. s7 (c)-(e)) in larger timescale. This directionality is hampered close to the topological defects in the VL. In the isotropic vortex liquid (IVL) phase the vortices move in random directions and show large jumps (fig. s7 (f)). Vortex creep persisting down to 450 mK is another proof that the liquid phases exist down to that temperature.

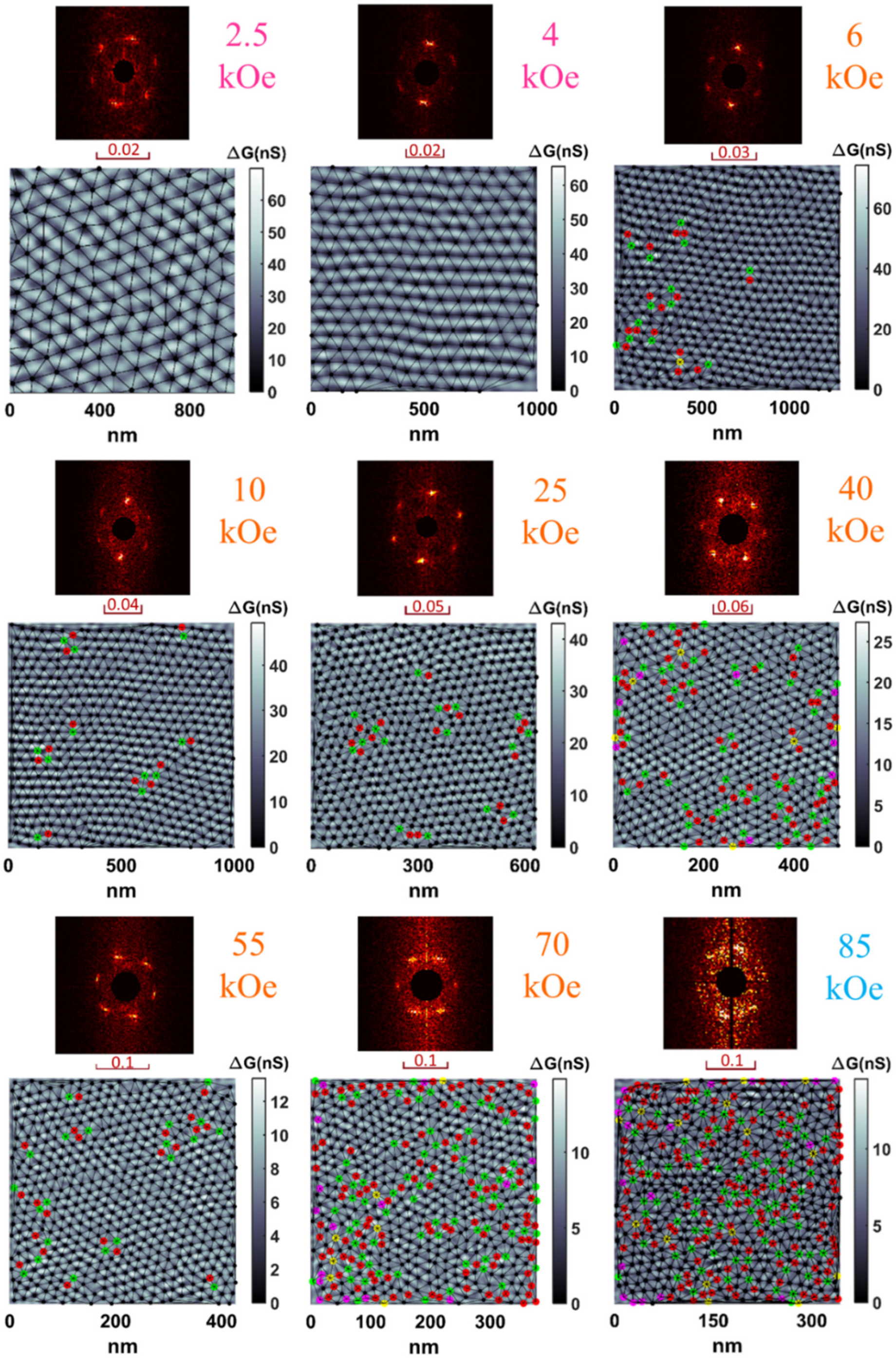

**Figure s6|** Vortex lattice images at 450 mK in different magnetic fields. Vortices shown as black dots correspond to the local minima of the conductance maps acquired at fixed bias voltage (1.45 mV) close to the coherence peak. The VL is Delaunay triangulated to find topological defects (denoted as red, green, magenta, and yellow dots) corresponding to five-, seven-, four-, and eightfold coordination. Above each vortex image is the 2D FT of image; the scale bar is in $nm^{-1}$.

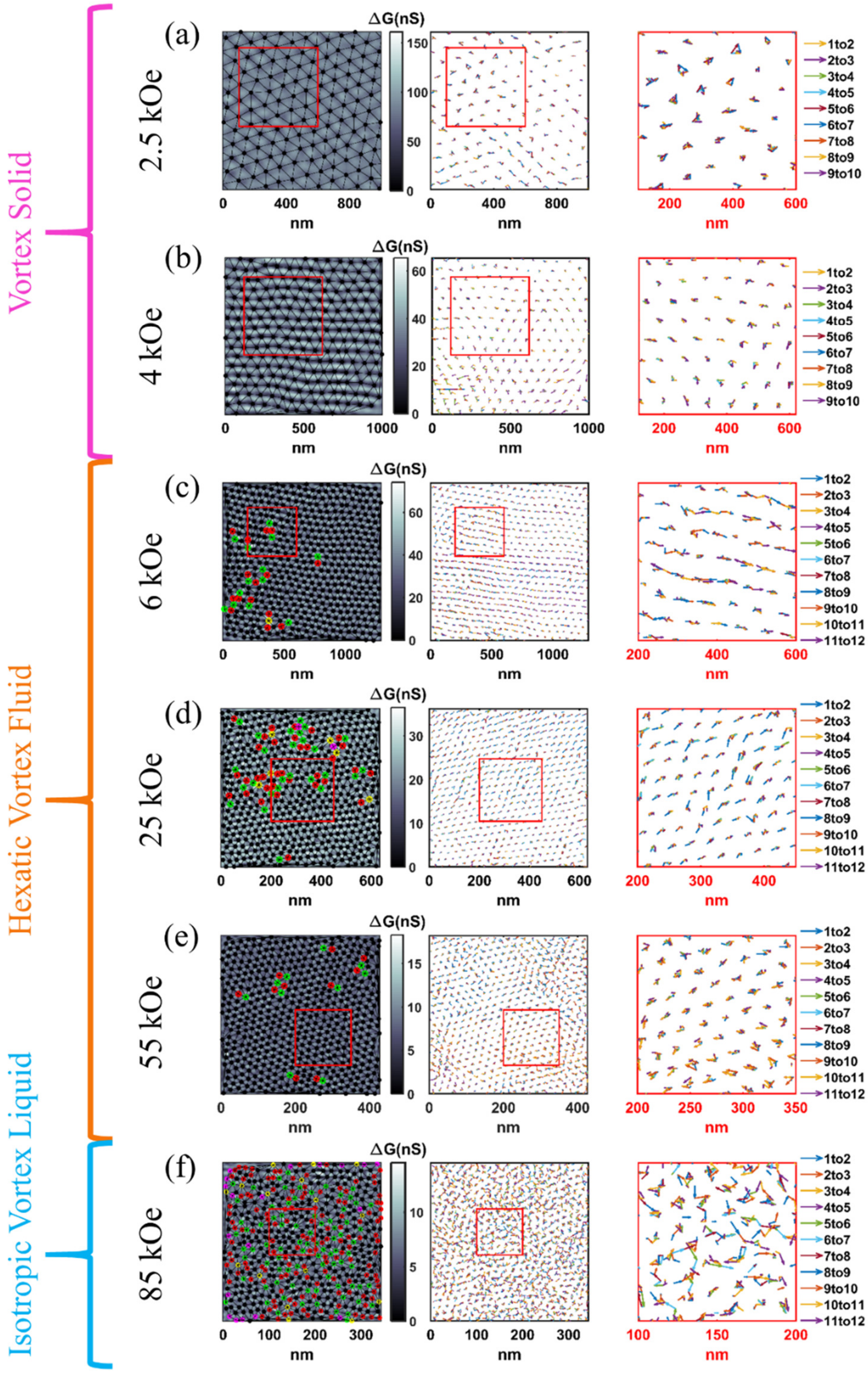

**Figure s7|** Vortex creep at 450 mK: (a)-(f) The leftmost panels show first vortex image of consecutive 10 or 12 images for different magnetic fields written left to each image. The black dots are positions of the vortices which are Delaunay triangulated to find topological defects denoted as red, green, magenta and yellow dots designating 5-, 7-, 4- and 8-fold coordination. The middle panels show arrows connecting consecutive 10 or 12 vortices. The red squares in these panels are expanded in the rightmost panels. Here vortices in (a)-(b) are in the Vortex solid regime, (c)-(e) are in Hexatic Vortex Fluid regime and (f) is in Isotropic Vortex Liquid regime.

### III. Simulating the conductance map for a vortex lattice

To cross-check the validity of the phenomenological model used to simulate the conductance maps for a vortex lattice, we apply it on a conventional system, namely the vortex state in clean 2H-$NbSe_2$ (single crystal), where we do not have any evidence that the vortices fluctuate about their mean positions. Fig. 8s(a) shows the VL image on a $NbSe_2$ single crystal at 5 kOe, 450 mK, keeping the bias voltage at 1.2 meV, close to the coherence peak. Fig. s8(b) shows the $G_N(V)$ vs $V$ spectra along a line passing through the center of the vortex. Here, the core of vortex shows a zero bias conductance peak ($G_N(V=0)>1$) resulting from the bound state of normal electrons inside the normal core, known as Caroli-de Gennes-Matricon bound state[3]. On the other hand, spectra obtained at superconducting regions away from the vortex core has regular BCS characteristics, partially broadened by the circulating current around the vortex core (*inset* Fig. s8(c)). The experimental variation of $G_N(0)$ along three lines passing through the center of the vortex as well as their average is shown in Fig. s8 (c).

The VL is simulated as follows. We assume that far away from the vortex core $G_N(V)$ will be similar to the conductance spectra in zero field. This is obtained by fitting the zero field experimental tunneling conductance with BCS theory using superconducting energy gap, $\Delta$ and phenomenological broadening parameter, $\Gamma$ as the fitting parameters; we call this resultant best fit spectrum as $G_N^{BCS}(V)$. For the spectra at the center at the vortex center, $G_N^{center}(V)$ we used the experimental spectra at the center of the vortex. These two spectra are shown in Fig. s8 (d). To interpolate between these two, we use an empirical Gaussian weight factor, $f(\boldsymbol{r})=\exp\left(-\frac{r^2}{2\sigma^2}\right)$, such that $G_N(V,\boldsymbol{r})=f(\boldsymbol{r})G_N^{center}(V)+[1-f(\boldsymbol{r})]G_N^{BCS}(V)$, where $\boldsymbol{r}$ is the position with respect to the center of the vortex. We choose $\sigma\approx 1.17\xi$ and where $\xi$ is the Ginzburg-Landau coherence length of clean $NbSe_2$ ($\xi_{GL}^{NbSe_2}\sim 8.9\ nm$, corresponding to the measured $H_{c2}\sim$ 42 kOe). Using this we construct the VL by linear superposition of conductance values from all vortices and henceforth obtain resultant conductance map given by, $\tilde{G}_N(V,\boldsymbol{r})=1.6*\sum_i G_N(V,\boldsymbol{r}-\boldsymbol{r_i})/[\sum_i G_N(V=0,\boldsymbol{r}-\boldsymbol{r_i})]_{max}$ where position of *i*-th vortex is $\boldsymbol{r}_i$. This construct ensures that at the center of the vortex the simulated conductance matches with the experimental zero bias conductance at the vortex center. In Fig. s8(e) we compare the experimental variation of $G_N(0)$ for a line passing through the vortex center along with $\widetilde{G}_N(0)$ obtained from our simulation. The *inset* shows the corresponding data for $G_N(1.2\ \text{meV})$ and $\widetilde{G}_N(1.2\ \text{meV})$ which is close to the superconducting coherence peaks. The good agreement in both cases shows the validity of our phenomenological approach to simulate the conductance map.

One difference between the above simulation and the one in the case in *a*-MoGe is that in that case we take $G_N^{center}(V) = 1$. The reason is that zero bias conductance peak at the center of the vortex core is very sensitive to disorder and very small amount of scattering (such as small amount of Co doping in $NbSe_2$) destroys the peak giving a flat spectrum[4]. Therefore it is unlikely to be present even in the absence of fluctuations in *a*-MoGe where the electronic mean free path is very small due to the amorphous nature of the sample.

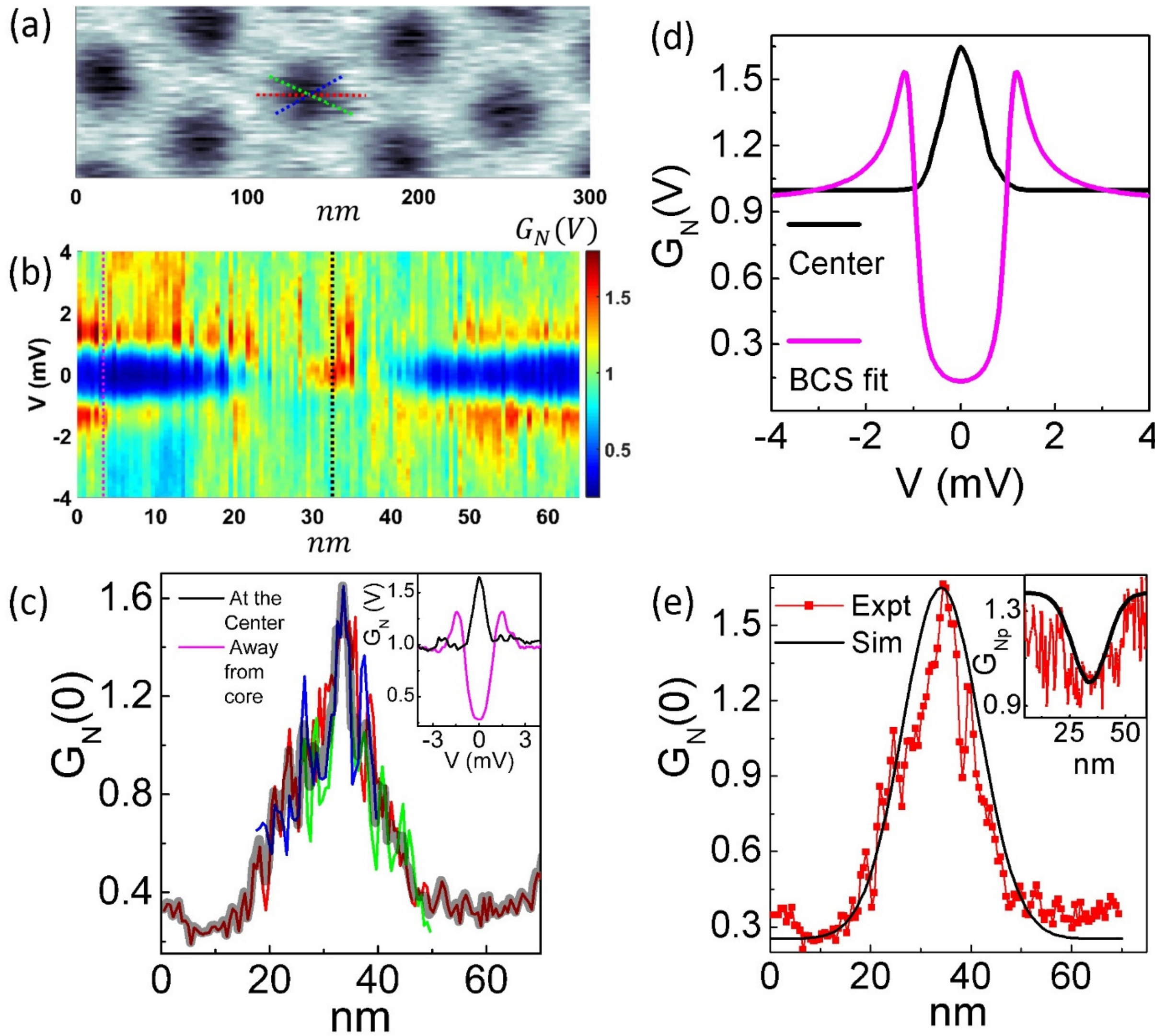

**Figure s8|** *Simulating vortex core in NbSe₂ for 5 kOe at 450 mK* (a) Vortex image, observed by recording $\Delta G(V = 1.2\ meV)$ across the area, showing three lines passing through a vortex center. (b) $G_N(V)$ vs. $V$ spectra along the red dotted line in (a) with black dotted line denoting the center of the vortex. (c) Variation of $G_N(0)$ along the dotted lines passing through the vortex cores in (a); *inset* Two spectra at the black and pink dotted lines in (b) respectively. The transparent grey is the average of the three. (d) The two spectra at the center of the vortex core and far away from the vortex core taken for the purpose of our simulation. (e) Simulated $\tilde{G}_N(0)$ (black line) along with the average $G_N(0)$ (red squares) for three lines passing through the center of the vortex obtained from experimental data. (*inset*) Simulated $\tilde{G}_N(V = 1.2\ meV)$ (black line) along with the average $G_N(V = 1.2\ meV)$ (red line) obtained from experimental data.

## IV. Harmonic approximation of vortex confining potential

First, we consider a unit cell of 2D vortex lattice (VL). Here, contribution of elastic energy in the VL comes in two ways: compression and shear. Between compression ($C_{11}$) and shear ($C_{66}$) elastic moduli, $C_{66} << C_{11}$, and therefore for small deformation the elastic energy of the VL is mainly controlled by $C_{66}$. Let's assume a single vortex is displaced by a distance δ (<< $a$, is lattice constant) from its equilibrium position (lattice point). The shear strain in the lattice,

$$\tan(\theta) \approx \theta = \frac{\delta}{a} \tag{1}$$

This causes elastic energy per unit volume,

$$\varepsilon = \frac{1}{2}\, C_{66} \left(\frac{\delta}{a}\right)^2 \tag{2}$$

So, the total elastic energy, $E = \varepsilon A d = \frac{1}{2}\, C_{66} \left(\frac{\delta}{a}\right)^2 Ad$, where A is area of unit cell and d is the thickness of film.

In the London limit, the shear modulus of an isotropic superconductor for small magnetic field is given by[5],

$$C_{66} = \frac{\emptyset_0 H}{16\pi\mu_0\lambda^2} \tag{3}$$

Using this expression, the total energy is,

$$E = \frac{\emptyset_0 d}{32\pi\mu_0\lambda^2}(HA)\left(\frac{\delta}{a}\right)^2 \tag{4}$$

Since we have one vortex per unit cell, $HA = \Phi_0$ ($\Phi_0$ is quantum flux).

So, total elastic energy,

$$E = \frac{{\emptyset_0}^2 d}{32\pi\mu_0\lambda^2}\left(\frac{\delta}{a}\right)^2 = \frac{1}{2}\left(\frac{{\emptyset_0}^2 d}{16\pi\mu_0\lambda^2 a^2}\right)\delta^2. \tag{5}$$

Therefore, motion of vortices about its equilibrium position, is looks like harmonic oscillator with frequency,

$$\omega = \sqrt{\frac{{\emptyset_0}^2 d}{16\pi\mu_0\lambda^2 a^2 m_v}} \tag{6}$$

Where, $m_v$ is vortex mass. The effective spring constant of this oscillator,

$$K_{eff} = \frac{\emptyset_0^2 d}{16\pi\mu_0\lambda^2}\frac{1}{a^2} = \frac{K_0}{16a^2}. \tag{7}$$

A more intuitive but less rigorous way to understand this result is by considering a 1D chain of vortices (fig. s9). This can be thought as any principal direction of VL. Interaction potential energy between vortex-vortex is given by[6],

$$U(x) = \frac{\emptyset_0^2 d}{2\pi\mu_0\lambda^2}\ln\left(\frac{\lambda}{x}\right) \tag{8}$$

Where $x$ is distance between two vortices ($\xi < x < \lambda$).

So the repulsive force felt by one vortex due to presence of another vortex at distance $x$ is given by,

$$f(x) = \frac{\emptyset_0^2 d}{2\pi\mu_0\lambda^2}\frac{1}{x} \tag{9}$$

Now if one vortex is displaced from its equilibrium position by a distance $\delta$ (<< $a$) ( shown in blue color) then the net restoring force on the displaced vortex is given by,

$$F(\delta) = \frac{\emptyset_0^2 d}{2\pi\mu_0\lambda^2}\sum_n\left(\frac{1}{na-\delta} - \frac{1}{na+\delta}\right) \approx \frac{\emptyset_0^2 d}{\pi\mu_0\lambda^2}\sum_n\left(\frac{\delta}{n^2a^2}\right) = \left(\frac{\pi^2}{6}\right)\left(\frac{\emptyset_0^2 d}{\pi\mu_0\lambda^2}\frac{1}{a^2}\right)\delta \tag{10}$$

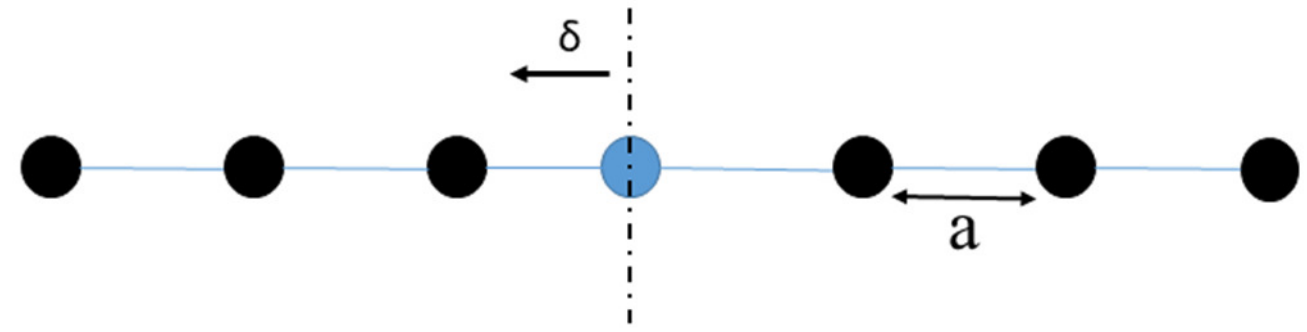

**Figure s9|** Schematic illustration of a 1D array of vortices with lattice spacing *a*. When a vortex is displaced by an amount δ from its equilibrium position, it experiences a restoring force due to the interaction with other vortices.

Where *a* is lattice constant. This net force follows Hook's law as expected. Therefore the effective potential is harmonic with spring constant,

$$K_{eff} = \left(\frac{\pi^2}{6}\right)\frac{\emptyset_0^2 d}{\pi\mu_0\lambda^2}\frac{1}{a^2} = \left(\frac{\pi^2}{6}\right)\frac{K_0}{a^2} \tag{11}$$

This expression has the same form as (7) except for the pre-factor.

From (7) we can also determine the resonant frequency of the oscillator,

$$\omega = \sqrt{\frac{\emptyset_0^2 d}{16\pi\mu_0\lambda^2}\frac{1}{m_v a^2}} = \omega_0\sqrt{h}$$

Where, $h = \frac{H}{Hc_2}$ and $\omega_0 = \frac{1}{1.075}\sqrt{\frac{\emptyset_0 d}{16\pi\mu_0\lambda^2}\frac{Hc_2}{m_v}}$ . Using $m_v = 35m_e$, as determined in the main paper, we obtain, $\omega_0 = 1.07 \times 10^{12}$ Hz.

**V. Fit of the $G_N(0)$ profile passing through the vortex core.** In Fig. 3(e)-3(f) we showed three representative fits of our model with $G_N(0)$ profiles passing through the vortex core. In Fig. s10 (a)-(h) we show the fits at several other magnetic fields, along with the best fit parameters. Fig. s10 (i) shows the value of Γ used to fit the profiles. Above 10 kOe we need to gradually increase the value of Γ to take into account the broadening arising from orbital current around the vortex core.

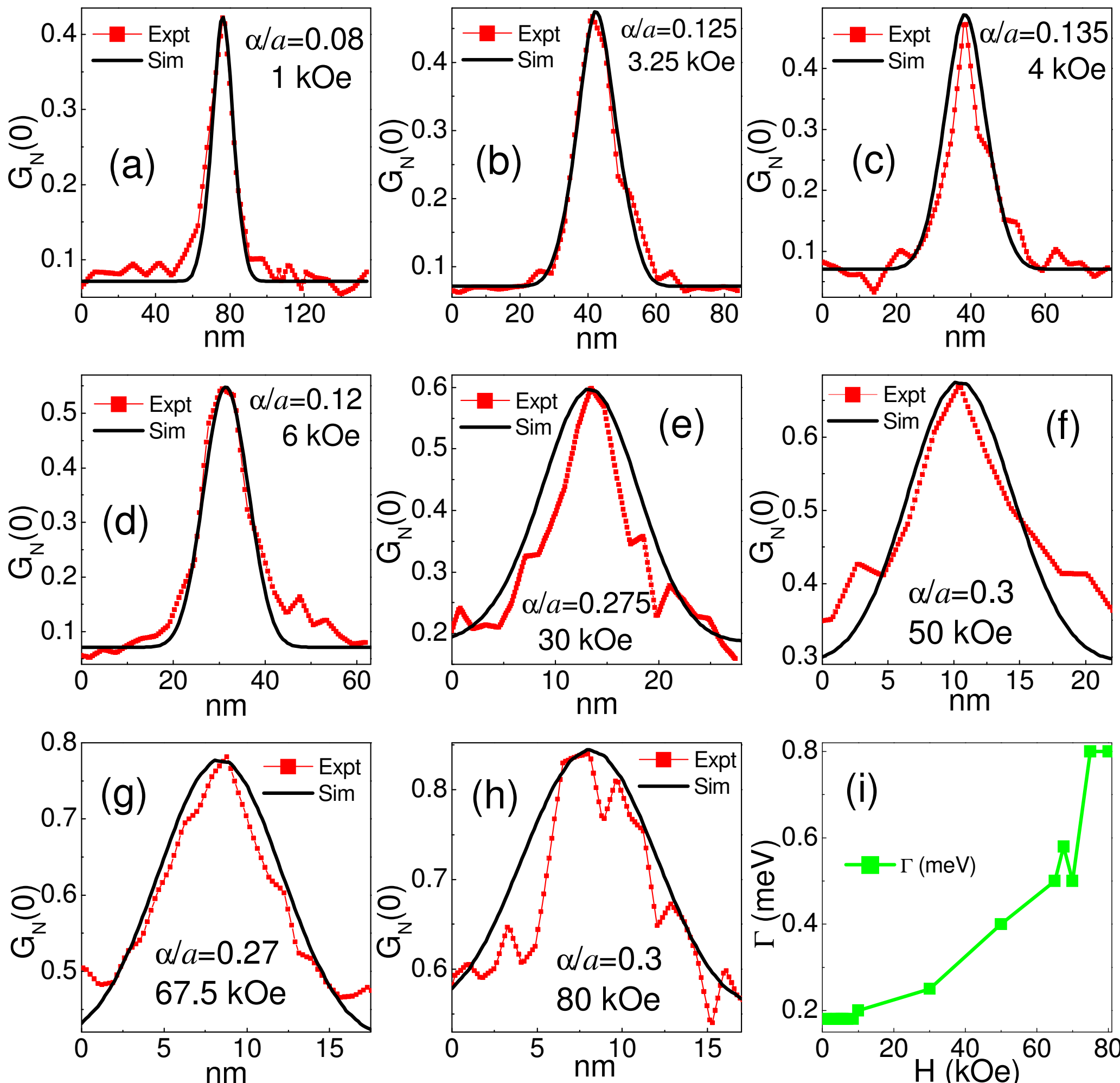

**Figure s10|** (a)-(g) The red connected points are averaged line-cuts of $G_N(0,r)$ along vortex cores in different magnetic fields. The black lines are line-cuts of simulated $\langle \tilde{G}_N(0,r)\rangle_\alpha$ of the corresponding field. (i) Magnetic field variation of the broadening parameter Γ used in the fits.

[1] I. Roy, S. Dutta, A. N. Roy Choudhury, Somak Basistha, Ilaria Maccari, Soumyajit Mandal, John Jesudasan, Vivas Bagwe, Claudio Castellani, Lara Benfatto, and Pratap Raychaudhuri, *Melting of the Vortex Lattice through Intermediate Hexatic Fluid in an a-MoGe Thin Film,* Phys. Rev. Lett. **122,** 047001 (2019).

[2] N. R. Werthamer, E. Helfland, and P. C. Honenberg, *Temperature and Purity Dependence of the Superconducting Critical Field,* $H_{c2}$*. III. Electron Spin and Spin-Orbit Effect,* Phys. Rev. **147**, 295 (1966).

[3] C. Caroli, P. G. de Gennes and J. Matricon, *Bound fermion states on a vortex line in a type II superconductor*, J. Phys. Lett. **9**, 307 (1964).

[4] S. C. Ganguli, H. Singh, R. Ganguly, V. Bagwe, A. Thamizhavel and P. Raychaudhuri, *Orientational coupling between the vortex lattice and the crystalline lattice in a weakly pinned* $Co_{0.0075}NbSe_2$ *single crystal*, J. Phys.: Condens. Matter **28**, 165701 (2016).

[5] E. H. Brandt, *The Vortex Lattice in Conventional and High-*$T_c$ *Superconductors,* Braz. J. Phys. **32**, 675 (2002).

[6] G. Blatter, M. V. Feigel'man, V. B. Geshkenbein, A. I. Larkin, and V. M. Vinokur, *Vortices in high-temperature superconductors,* Rev. Mod. Phys. **66**, 1125 (1994).